\documentclass[final,5p,times,twocolumn,authoryear]{elsarticle}
\usepackage{subfigure}
\usepackage{amssymb}
\usepackage{CJK}
\usepackage{amsmath}
\usepackage{graphicx}

\usepackage {threeparttable}
\journal{High Energy Astrophysics}

\begin{document}

\begin{frontmatter}

\title{Constraints on the internal physics of neutron stars from the observational data of several young pulsars: the role of a power-law decaying dipole magnetic field}

\author{Yu-Long Yan}
\author{Quan Cheng$^\ast$}
\ead{qcheng@ccnu.edu.cn}
\author{Xiao-Ping Zheng$^\dagger$}
\ead{zhxp@ccnu.edu.cn}
\affiliation{organization={Institute of Astrophysics},
            addressline={Central China Normal
University}, 
            city={Wuhan},
            postcode={430079}, 
            country={China}}

\begin{abstract}
The observational data (e.g., the timing data and magnetic tilt angles $\chi$) of young pulsars can be used to probe some critical issues about the internal physics of neutron stars (NSs), for instance, the number of precession cycles $\xi$ and the internal magnetic field configuration (IMFC) of NSs. Evolution of the dipole magnetic field $B_{\rm d}$ of NSs may play an important role in determining the final results. In this work, a power-law form is adopted to describe the decay of $B_{\rm d}$. In such a scenario, the IMFC and $\xi$ of young pulsars with an ordinary $B_{\rm d}\sim10^{12}-\-10^{13}$ G and a steady braking index $n$ are investigated. Since the tilt angle change rates $\dot{\chi}$ of pulsars with $n<3$ can be theoretically predicted, a test on the power-law decay model can thus be made by comparing the theoretical values to that obtained from observations. However, such a comparison can only be made on the Crab pulsar currently, and the results show that the power-law decay model is inconsistent with the Crab's observations. We suggest that rather than decay, the Crab's $B_{\rm d}$ should increase with time at a rate $\sim12-14$ G/s. A definite conclusion on the validity of the power-law decay model for pulsars with ordinary $B_{\rm d}$ may be given if $\dot{\chi}$ of other pulsars could be measured.
\end{abstract}

\begin{keyword}
neutron \sep magnetic fields \sep pulsars \sep graviational waves \sep superfluid
\end{keyword}

\end{frontmatter}




\section{Introduction}
\label{sec:intro}
The braking indices of pulsars represent the spin-down behavior of neutron stars (NSs) and can be associated with various aspects of NS physics, for instance, the physics in the outer magnetosphere of NSs \citep{2001ApJ...561L..85X,2006ApJ...643.1139C,2015MNRAS.450.1990K,2021ApJ...920...57C}, the dense matter equation of state \citep{2012NatPh...8..787H}, the superfluid physics of dense matter \citep{2012NatPh...8..787H}, and the magnetic fields of NSs \citep{2015MNRAS.452..845H,2015MNRAS.446.1121G,2019PhRvD..99h3011C}. As a result, the braking indices measured from timing observations of pulsars provide us a vital probe to investigate NS physics in addition to observations of surface thermal radiation of NSs \citep{2006NuPhA.777..497P,2021MNRAS.506..709S,2022MNRAS.512.4689H} and gravitational waves from binary NS mergers \citep{2017PhRvL.119p1101A,2023PhRvL.130i1404F}. 

Until now, the braking indices of only nine young pulsars have been precisely measured via radio and X-ray timing observations \citep{2015MNRAS.446..857L,2016ApJ...819L..16A}. It is remarkable that all these young pulsars have braking indices $n\neq3$ and only one pulsar (PSR J1640-4631) has $n>3$ \citep{2016ApJ...819L..16A}. The diversity of the values of $n$ suggest that the classical magnetic dipole (MD) model cannot properly account for the observational results and thus other mechanisms are needed. A combination of the classical MD radiation and other braking mechanisms thus have been suggested in lots of theoretical work to interpret the observed braking indices (e.g., \citealt{2001ApJ...554L..63M,2001ApJ...561L..85X,2012NatPh...8..787H,2015MNRAS.452..845H,2015MNRAS.446.1121G,2016ApJ...823...34E,2017ApJ...837..117T,2016A&A...593L...3C,2017ApJ...849...19G}). 

Theoretically, since the NSs not only have strong external dipole magnetic fields but also probably possess even stronger internal magnetic fields consist of both the poloidal and toroidal components \citep{2004Natur.431..819B,2006A&A...450.1097B,2009MNRAS.397..763B}, the gravitational wave (GW) radiation from quadruple deformation caused by the strong internal fields will naturally contribute to the measured $n$ \citep{2019PhRvD..99h3011C,2023RAA....23e5020H}. Moreover, for a magnetically deformed NS, free-body precession of the star's magnetic axis around spin axis will take place if the NS is not in the minimum spin-energy state \citep{2009MNRAS.398.1869D}. The precessional energy of the NS can be dissipated due to internal viscosity and thus the magnetic tilt angle between the spin and magnetic axes can evolve with time \citep{2009MNRAS.398.1869D,2019PhRvD..99h3011C}. Evolution of the tilt angle on one hand depends on the shape of the magnetically deformed NS \citep{2001MNRAS.324..811J,2009MNRAS.398.1869D,2019PhRvD..99h3011C}, on the other hand relies on a critical physical parameter called the number of precession cycles, $\xi$ \citep{2002PhRvD..66h4025C,2009MNRAS.398.1869D,2019PhRvD..99h3011C}. The shape of the deformed NS is actually dependent on the structure of its internal magnetic fields. The deformed NS will be an oblate ellipsoid ($\epsilon_{\rm B}>0$) if its internal magnetic fields are dominated by the poloidal component, whereas it will be a prolate ellipsoid ($\epsilon_{\rm B}<0$) if the toroidal part plays a dominant role in the internal fields. Damping of the free-body precession of the NS due to internal viscosity will cause decrease of the tilt angle if the NS has poloidal-dominated internal fields \citep{2002PhRvD..66h4025C,2009MNRAS.398.1869D,2019PhRvD..99h3011C}. On the contrary, in the case of toroidal-dominated internal fields, damping the free-body precession due to viscosity can increase the NS's tilt angle \citep{2002PhRvD..66h4025C,2009MNRAS.398.1869D,2019PhRvD..99h3011C}. The changing rate of the tilt angle, however, is related to $\xi$, whose value is still under debate \citep{2002PhRvD..66h4025C,2019PhRvD..99h3011C}. This parameter indicates the specific viscous mechanisms through which the precessional energy of the NS is dissipated, on the other hand represents the interactions between superfluid neutrons and other particles in the NS interior \citep{2009MNRAS.398.1869D,2019PhRvD..99h3011C}. Consequently, an investigation on $\xi$ may not only be necessary for the study of continuous GW radiation from single NSs but also be extremely important for the research of complex interactions between superfluid neutrons and other particles \citep{2019PhRvD..99h3011C,2023RAA....23e5020H}. It should also be noted that MD and GW radiation can lead to aligned torques between the spin and magnetic axes and hence affect the tilt angle evolution \citep{2000PhRvD..63b4002C,2009MNRAS.398.1869D,2014MNRAS.441.1879P}. In short, with the presence of strong magnetic fields of NSs, the tilt angle evolution may be a natural consequence and its effect on the braking indices should be involved in theoretical calculations. 

As the spin-down of NSs, their dipole magnetic fields may decay with time due to the combined effects of Hall drift and Ohmic dissipation because the resistance of NS crusts is probably non-zero \citep{1992ApJ...395..250G,2008A&A...486..255A,2020MNRAS.494.3790K}. The effect of dipole field decay on the measured braking indices of pulsars has been considered in a lot of work previously (e.g., \citealt{2001A&A...376..543T,2012ApJ...757..153Z,2017ApJ...849...19G,2019PhRvD..99h3011C}). Though the consensus is that both Hall drift and Ohmic dissipation may play an important role in the field decay if the dipole fields indeed stem from NS crusts, the mathematical form of field decay is still uncertain \citep{2019PhRvD..99h3011C}. In previous work, it has been suggested that decay of the dipole field may follow an exponential, a nonlinear, or even a power-law form \citep{2017ApJ...849...19G}. By adopting the exponential and nonlinear decay forms of dipole field, and using the observational data of young pulsars with a steady braking index, \cite{2019PhRvD..99h3011C} and \cite{2023RAA....23e5020H} have set constraints on the number of precession cycles and the structure of internal magnetic fields of these pulsars. 

In this paper, assuming that the dipole-field decay follows the power-law form as that suggested in \cite{2000ApJ...529L..29C}, and using the observational data of young pulsars with a steady braking index and an ordinary dipole field, we first set constraints on $\xi$ and the internal magnetic field structure of these pulsars. We find that though a different form for the dipole-field decay is adopted, in most cases the constraints on $\xi$ and the internal field structure remain unchanged in comparison with the results obtained based on the form of exponential decay \citep{2019PhRvD..99h3011C,2023RAA....23e5020H}. Specifically, our results show that in the power-law decay scenario, without measurements of the tilt angle of PSR J1640-4631 (which has $n>3$), we cannot deduce whether its internal magnetic fields are dominated by the poloidal or the toroidal component if its dipole field decays at a very small rate. However, for young pulsars with $n<3$, their internal fields are probably dominated by the toroidal parts. From the measured values of the tilt angles of these pulsars, the range of the number of precession cycles is generally constrained to be $10^4\lesssim\xi\lesssim{\rm a~few}\times10^6$, which is consistent with the constraints obtained by assuming the exponential decay of the dipole field \citep{2023RAA....23e5020H} and also that derived from modeling of the glitch rise behaviors of the Crab and Vela pulsars \citep{2018MNRAS.481L.146H,2019NatAs...3.1143A}.\footnote{The parameter $\xi$ is approximately equal to the reciprocal of the superfluid mutual friction parameter, $\mathcal{B}$, whose value can be constrained by modeling of the glitch rise processes of some pulsars \citep{2018MNRAS.481L.146H,2019NatAs...3.1143A}.} 

Depending on the values of the parameters, possible physical mechanisms behind the power-law form of dipole-field decay proposed in \cite{2000ApJ...529L..29C} are ambipolar diffusion of the core magnetic fields in the irrotational or solenoidal mode, or even Hall cascade of the crustal fields. It should be noted that though some previous studies showed that both ambipolar diffusion in the core and Hall drift in the crust may not lead to magnetic field decay after the field reached an equilibrium configuration \citep{2014PhRvL.112q1101G,2020MNRAS.498.3000C}, more elaborate three-dimensional simulations and observational evidences are needed before giving a definite conclusion on this issue \citep{2015MNRAS.447.1213M,2017JPhCS.932a2048P,2019MNRAS.490.2013C}. The power-law form of dipole-field decay may only apply to highly magnetized NSs with dipole magnetic fields $B_{\rm d}>10^{13}$ G \citep{2000ApJ...529L..29C}, though an extrapolation of the decay form to NSs with $B_{\rm d}\sim10^{11}-\-10^{13}$ G was performed in \cite{2001SHEP...15..381P,2001A&AT...20..635P} to account for some X-ray sources observed by ROSAT. As a result, for NSs with dipole fields of ordinary strength ($B_{\rm d}\sim10^{12}-\-10^{13}$ G), a test of the validity of the power-law decay model is necessary, especially when considering that the mathematical form of field decay remains mysterious. This represents one of the main purposes of our work. Since the change rate of tilt angle can be derived from our model, such a test can be performed by comparing the theoretically calculated tilt angle change rate to that obtained from observations. Such a test, however, can only be performed on the Crab pulsar because it is the only pulsar whose tilt angle change rate can be observationally derived to date \citep{2013Sci...342..598L}. Our results show that in the power-law decay model, the calculated tilt angle change rate is several times larger than the inferred vale from observations. Moreover, if the dipole field of the Crab pulsar indeed decays following the power-law form, in order to satisfy all of the Crab's timing data and the inferred tilt angle change rate, an unreasonable large $\xi\sim10^9$ is required. Therefore, in view of the two problems, observations of the Crab pulsar may exclude the power-law decay model. A definite conclusion on the validity of the power-law decay model may be given if the tilt angle change rate of other young pulsars can be measured in future observations. Finally, we find that to be in line with all observations of the Crab pulsar, rather than decay with time, its dipole field should increase at a rate $\sim12-\-14$ G/s. Meanwhile, the number of precession cycles is derived to be within the range $1.08\times10^4\lesssim\xi\lesssim2.43\times10^5$, generally consistent with the results obtained from modeling of the glitch rise processes of the Crab pulsar \citep{2018MNRAS.481L.146H}.

This paper is organized as follows. We introduce the model for calculating the braking indices of pulsars and the power-law decay model of dipole fields in Section \ref{model}. The derived constraints on $\xi$ and the structure of internal magnetic fields in the power-law decay model are presented in Section \ref{results}. In this section, we also test the power-law decay model with observations of the Crab pulsar, and set new constraints on $\xi$ and obtain the dipole-field increase rate for the Crab. Finally, conclusion and discussions are given in Section \ref{conclusion}.

\section{Evolution of pulsars}
\label{model}
As generally considered, evolution of the spin periods, tilt angles, and magnetic fields of pulsars can all contribute to their measured braking indices. Because of the presence of strong external and internal magnetic fields, the NSs can spin down through both MD and GW radiation \citep{1983bhwd.book.....S}. Meanwhile, the tilt angles of the magnetically deformed NSs can evolve with time due to MD and GW radation, and damping of the free-body precession of the stars induced by internal viscous dissipation \citep{2009MNRAS.398.1869D,2019PhRvD..99h3011C}. The dipole magnetic fields of the NSs may decay with time due to the combined effects of Hall drift and Ohmic decay if they stem from the NS crusts \citep{1992ApJ...395..250G,2004ApJ...609..999C,2008A&A...486..255A,2020MNRAS.494.3790K}. However, if the dipole fields follow the change of the core fields of NSs, as the latter may decay because of ambipolar diffusion \citep{1992ApJ...395..250G,1998ApJ...506L..61H,2016ApJ...833..261B}, the former may possibly decay on the same timescale. After taking into account the spin-down, tilt angle, and magnetic field evolutions, and noticing that the pulsars may probably be surrounded by a plasma-filled magnetosphere \citep{1969ApJ...157..869G,2006ApJ...648L..51S}, the expression for the braking index can be theoretically derived as \citep{2019PhRvD..99h3011C,2023RAA....23e5020H}
\begin{equation}\label{bi}
\begin{aligned}
n=& 3-\frac{2 P}{\dot{P}}\left\{\frac{\dot{B}_{\mathrm{d}}}{B_{\mathrm{d}}}+\dot{\chi} \sin \chi \cos \chi\left[\frac{1}{1+\sin ^{2} \chi}\right.\right.\\
&\left.\left.+\frac{1+30 \sin ^{2} \chi}{\eta \sin ^{2} \chi\left(1+15 \sin ^{2} \chi\right)}\right]\right\},
\end{aligned}
\end{equation}
where $P$, $B_{\rm d}$, and $\chi$ are respectively the spin
period, dipole magnetic field, and magnetic tilt angle of a pulsar.
$\dot{P}$, $\dot{B}_{\rm d}$, and $\dot{\chi}$ are the first
derivatives of the above quantities. $\eta=5kc^2B_{\rm
d}^2R^6(1+\sin^2\chi)/[2G\epsilon_{\rm
B}^2I^2\Omega^2(1+15\sin^2\chi)\sin^2\chi]$ represents the ratio of
MD spin-down to GW spin-down rates with $R$, $I$, $\epsilon_{\rm
B}$, and $\Omega=2\pi/P$ denoting the radius, moment of inertia,
quadruple ellipticity of magnetic deformation, and angular frequency
of the NS, respectively. $k$ is the coefficient related to MD
radiation. In this work, we take $k=1/6$, $R=10$ km, and
$I=10^{45}~{\rm g}~{\rm cm}^2$ \citep{2019PhRvD..99h3011C}. After neglecting the GW spin-down, the dipole field can be expressed as a function of $\chi$ via \citep{2019PhRvD..99h3011C,2023RAA....23e5020H}
\begin{equation}\label{Bd}
B_{\rm d}=\left[-\frac{P\dot{P}Ic^3}{4\pi^2kR^6(1+{\rm
sin}^2\chi)}\right]^{1/2}.   
\end{equation}
 
The internal magnetic field configuration of NSs plays an important role in determining the ellipticity of magnetic deformation. Although Numerical simulations showed that the internal fields may have a twisted-torus structure consists of both poloidal and toroidal fields \citep{2004Natur.431..819B,2006A&A...450.1097B}, the dominant component is still under debate. This issue is tightly related to the shape of the NS, thus can affect the tilt angle evolution, as stated in Section \ref{sec:intro}. In this paper, two expressions respectively correspond to the ellipticities of the poloidal-dominated (PD) and toroidal-dominated (TD) internal fields, are taken as $\epsilon_{\rm B}\simeq3.4\times10^{-7}(B_{\rm d}/10^{13}~{\rm G})$ and $\epsilon_{\rm B}\simeq-10^{-8}(\bar{B}_{\rm in}/10^{13}~{\rm G})$, where $\bar{B}_{\rm in}$ denotes the volume-averaged strength of the internal toroidal field \citep{2019PhRvD..99h3011C}. For a detailed description of the expressions one can refer to \cite{2019PhRvD..99h3011C}. Considering that the magnetically deformed NS may be embedded in a plasma magnetosphere, its tilt angle change rate has the following form \citep{2019PhRvD..99h3011C}:
\begin{equation}\label{chidot}
\dot{\chi}=\left\{\begin{array}{l}
-\frac{2 G}{5 c^{5}} I \epsilon_{\mathrm{B}}^{2} \omega^{4} \sin \chi \cos \chi\left(15 \sin ^{2} \chi+1\right)-\frac{\epsilon_{\mathrm{B}}}{\xi P} \tan \chi \\
-\frac{k B_{\mathrm{d}}^{2} R^{6} \omega^{2}}{I c^{3}} \sin \chi \cos \chi, \quad \text { for } \epsilon_{\mathrm{B}}>0 \\
-\frac{2 G}{5 c^{5}} I \epsilon_{\mathrm{B}}^{2} \omega^{4} \sin \chi \cos \chi\left(15 \sin ^{2} \chi+1\right)-\frac{\epsilon_{\mathrm{B}}}{\xi P} \cot \chi \\
-\frac{k B_{\mathrm{d}}^{2} R^{6} \omega^{2}}{I c^{3}} \sin \chi \cos \chi, \quad \text { for } \epsilon_{\mathrm{B}}<0 .
\end{array}\right.   
\end{equation}
The first and second terms of the above equation respectively represent the decrease of $\chi$ due to GW and MD radiation. The third term represents the change of $\chi$ caused by viscous damping of the NS's free-body precession, and one can see that the specific form of this term actually depends on the sign of $\epsilon_{\rm B}$, as stated in Section \ref{sec:intro}. 

The dipole field decay rate $\dot{B}_{\rm d}$ is determined by the specific field decay mechanisms. Although it is proposed that the magnetic fields in the core of a NS may decay because of ambipolar diffusion \citep{1992ApJ...395..250G,2016ApJ...833..261B}, while the fields in the crust may decay due to the combined effects of Hall drift and Ohmic dissipation \citep{1992ApJ...395..250G,2004ApJ...609..999C}, it is still unknown that whether the surface dipole field of a young pulsar decays with time \citep{2006ApJ...643..332F}, and if so, what mathematical form the dipole field decay will follow. The critical issue may be that whether decay of the dipole field follows decay of the crustal fields or the core fields. If decay of the dipole field follows decay of the crustal fields, it may also decay due to Hall drift and Ohmic dissipation, as investigated in \cite{2019PhRvD..99h3011C} and \cite{2023RAA....23e5020H}. Alternatively, if decay of the dipole field keeps in step with decay of the core fields, then its decay may be controlled by ambipolar diffusion. Previously, theoretical work suggested that dissipation of the core fields due to ambipolar diffusion and the crustal fields due to Hall cascade could both explain the high surface temperatures and relatively large spin periods of magnetars \citep{1998ApJ...506L..61H,2000ASPC..202..681G,2000ApJ...529L..29C}. Correspondingly, in these work, the dipole field was considered to decrease with time due to either ambipolar diffusion or Hall cascade. Although the dissipation mechanisms are different, the same power-law form, however, with different parameters can be used to describe decay of the dipole field. Following \cite{2000ApJ...529L..29C}, evolution of the dipole field reads
\begin{equation}\label{dBdt}
\frac{d B_{\rm d}}{d t}=-a B_{\rm d}^{1+\alpha},    
\end{equation}
where the parameters $a$ and $\alpha$ depend on specific dissipation mechanisms. Decay of $B_{\rm d}$ because of ambipolar diffusion can proceed in two modes: the irrotational mode and solenoidal mode \citep{1992ApJ...395..250G}. For the former mode, one has $a=0.01$ and $\alpha=5/4$, while for the latter mode, one has $a=0.15$ and $\alpha=5/4$ \citep{2000ApJ...529L..29C}. If decay of $B_{\rm d}$ is determined by Hall cascade, one has $a=10$, $\alpha=1$ \citep{2000ApJ...529L..29C}. \textbf{It should be noted that when deriving the three sets of values for $a$ and $\alpha$, $B_{\rm d}$ in unit of $10^{13}$ G and $t$ in unit of $10^6$ yrs are adopted (see \citealt{2000ApJ...529L..29C}).} Though the power-law form of field decay may hold for $B_{\rm d}>10^{13}$ G \citep{2000ApJ...529L..29C}, an extrapolation of the form to pulsars with $B_{\rm d}\sim10^{11}-\-10^{13}$ G was performed in \cite{2001SHEP...15..381P, 2001A&AT...20..635P}. Hence, the most important issue is that whether such a power-law decay form can be applied to pulsars with ordinary $B_{\rm d}$. For these pulsars, a test of the validity of power-law decay model using their radio observational data is thus necessary because it may not only justify or exclude the suggestion of power-law decay for $B_{\rm d}\sim10^{12}-\-10^{13}$ G, but also provide some clues on the field decay mechanisms. As we will show later, such a test can be performed by comparing the theoretically calculated $\dot{\chi}$ to that obtained from observations of pulsars. However, the test can only be done on the Crab pulsar since it is the only pulsar whose $\dot{\chi}$ has been inferred to date \citep{2013Sci...342..598L}.

Before initiating the theoretical calculations, it is interesting to discuss the consequences of power-law decay for young pulsars with $B_{\rm d}$ of ordinary strength. The observational data of young pulsars focused in this paper is summarized in Table \ref{tab:tab1}. These young pulsars all have a steadily measured braking index and a dipole field in the range $\sim10^{12}-\-10^{13}$ G. Moreover, the ages\footnote{Here we stress that the age of the Crab pulsar refers to its actual age $\tau_{\rm a}$, while the ages of other pulsars are their characteristic ages $\tau_{\rm c}=P/2\dot{P}$ because their actual ages are not available currently. See Table \ref{tab:tab1} for details.} of all these young pulsars are $\tau\lesssim10^4$ yrs. Hence, for the young pulsars at their current ages, if their dipole fields decay following the power-law form given in Equation (\ref{dBdt}), \textbf{for the sets of values of ($a$, $\alpha$), in most cases the decay rates satisfy $\dot{B}_{\rm d}\simeq0$}. This conclusion can be inferred from Figure 2 of \cite{2000ApJ...529L..29C} and Figure 1 of \cite{2001A&AT...20..635P}, which show that for young pulsars with $B_{\rm d}\sim10^{12}-\-10^{13}$ G and $\tau\lesssim10^4$ yrs, their dipole fields remain nearly constant in most cases if they follow the power-law decay form. \textbf{For each source, by adopting $0<\chi\leq\pi/2$, we can derive the range of $B_{\rm d}$ from Equation (\ref{Bd}), and further the ranges of $\dot{B}_{\rm d}$ through Equation (\ref{dBdt}) for the three sets of values of ($a$, $\alpha$). The results are shown in Table \ref{tab:tab2}, which suggests that generally the field decay rates are so small that we can approximately take $\dot{B}_{\rm d}\simeq0$ except for the only two cases, ($a$, $\alpha$)=(10, 1) for PSRs J1640-4631 and J1513-5908. Consequently, for the two cases, we will substitute $\dot{B}_{\rm d}$ obtained into Equation (\ref{bi}) to derive the curve of $\xi$ versus $\chi$. For other cases, when solving Equation (\ref{bi}) we can set $\dot{B}_{\rm d}\simeq0$ for simplicity. As an example, the effects of different $\dot{B}_{\rm d}$ on the final results are respectively shown in Figures \ref{fig1} and \ref{fig2} for PSRs J1640-4631 and J1513-5908.} 

\begin{table*}[h!]
	\centering
	\caption{The spin period $P$, its first derivative $\dot{P}$, age $\tau$, and measured tilt angle $\chi$ of young pulsars with a steady braking index $n$ measured and a dipole field in the range $\sim10^{12}-\-10^{13}$ G.}
        \label{tab:tab1}
        \begin{threeparttable}
        \setlength{\tabcolsep}{1.2pt}
	\begin{tabular}{ccccccc} 
		\hline
		{Pulsar name} & {$P$ (s)} & {$\dot{P}$ 
            $(10^{-13}~\rm{s/s})$} & {$\tau$ (yrs)} & {$\chi$} & {$n$} & {References}\\
		\hline
		PSR B0833-45 (Vela) & 0.089 & 1.25  & 11288 & $62^{\circ}$, 
            $70^{\circ}$, $75^{\circ}$, $79^{\circ}$ & $1.4\pm0.2$ & (1), (2), (3), (4) \\
		PSR J1833-1034 & 0.062 & 2.02 & 4866 & $70^{\circ}$ & 
            $1.8569\pm0.0006$ & (5), (6) \\
		PSR B0531+21 (Crab) & 0.033 & 4.21 & 969 & $45^{\circ}$,  
            $60^{\circ}$, $70^{\circ}$ & $2.51\pm0.01$ & (2), (3), (7), (8), (9) \\
            PSR J1513-5908 & 0.151 & 15.3 & 1564 & $3^{\circ}$, $10^{\circ}$ & $2.839\pm0.001$ & (10), (11) \\
            PSR J1640-4631 & 0.207 & 9.72 & 3376 & $-\-$ & $3.15 \pm 0.03$ & (12)\\
		\hline
	\end{tabular}
        \begin{tablenotes}
        \item{Reference.(1) \cite{1996Natur.381..497L}, (2) \cite{2009ApJ...695.1289W}, (3) \cite{2003ApJ...598.1201D}, (4) \cite{2016ApJ...832..107B}, (5) \cite{2012MNRAS.424.2213R}, (6) \cite{2013ApJ...765..124L}, (7) \cite{1993MNRAS.265.1003L}, (8) \cite{2008ApJ...680.1378H}, (9) \cite{2012ApJ...748...84D}, (10) \cite{2017ARep...61..591N}, (11) \cite{2007ApSS.308..317L}, (12) \cite{2016ApJ...819L..16A}.}
        \end{tablenotes}
\end{threeparttable}
\end{table*}

\section{Results} \label{results}
\subsection{Results for the power-law decay model}
\begin{figure*}[h!]
\centering
\subfigure[]{
\includegraphics[width=0.44\linewidth]{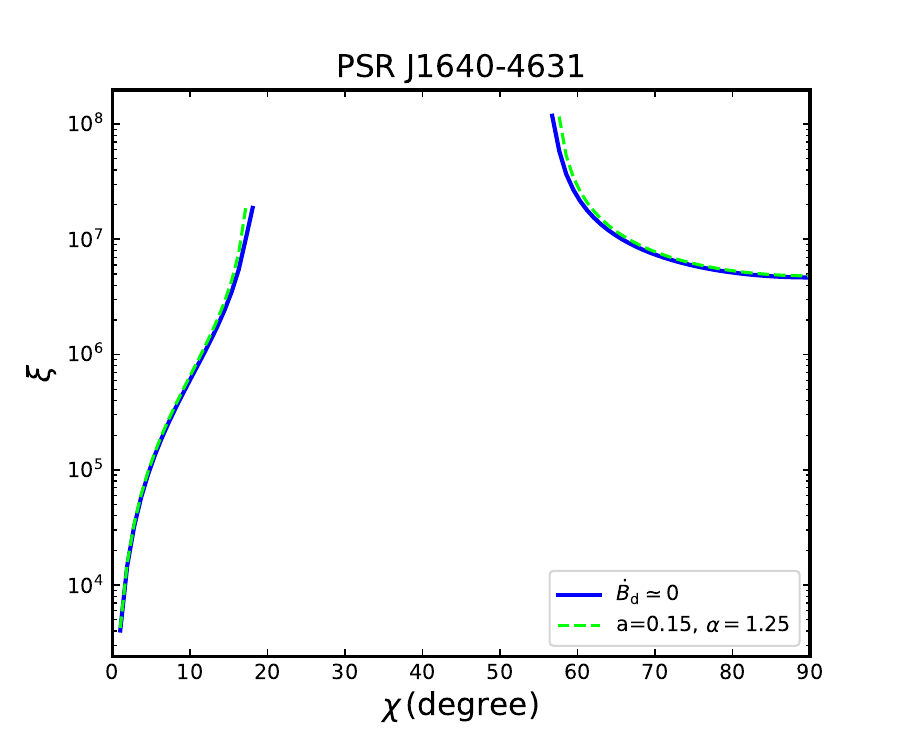}}
\hspace{0.01\linewidth}
\subfigure[]{
\includegraphics[width=0.44\linewidth]{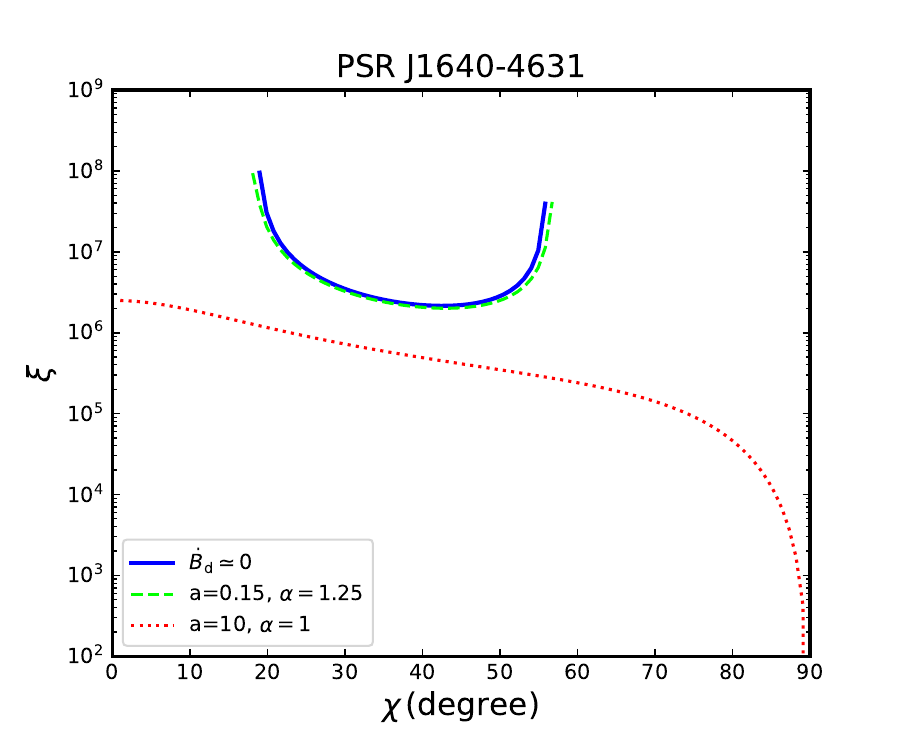}}
\caption{\label{fig1}{\small The number of precession cycles $\xi$ versus tilt angle $\chi$ calculated by using the measured $P$, $\dot{P}$, and $n$ of PSR J1640-4631. Panel (a) shows the results of PD internal fields, while panel (b) corresponds to the case of TD internal fields. The blue solid lines show the results of $\dot{B}_{\rm d}\simeq0$, while the green dashed, and red dotted lines respectively correspond to the results obtained by using the power-law decay form given in Equation (\ref{dBdt}) with ($a$, $\alpha$)=(0.15, 5/4), and (10, 1), as shown in the legends.}}
\end{figure*}

After substituting Equations (\ref{Bd}) and (\ref{chidot}), the values of $\dot{B}_{\rm d}$ determined above, and the measured $P$, $\dot{P}$ and $n$ of young pulsars into Equation (\ref{bi}), one can obtain the curve of $\xi$ versus $\chi$. In the case of TD internal fields, we take $\bar{B}_{\rm in}=10B_{\rm d}$. Moreover, the error bars in the values of $n$ are neglected in the calculations for simplicity \citep{2019PhRvD..99h3011C,2023RAA....23e5020H}. Before showing the quantitative results, from the measured timing data of these young pulsars, we may set some constraints on their internal magnetic field configurations, as that done in \cite{2023RAA....23e5020H}. For pulsars with $n<3$, if their internal fields are PD (i.e., $\epsilon_{\rm B}>0$), from Equation (\ref{chidot}) one can obtain $\dot{\chi}<0$. By substituting the value of $\dot{\chi}$ into Equation (\ref{bi}), we have $n>3$, which is inconsistent with observations. However, if the internal fields of pulsars with $n<3$ are TD (i.e., $\epsilon_{\rm B}<0$), depending on the value of $\xi$, from Equation (\ref{chidot}) we may have $\dot{\chi}>0$. As a result, the resultant braking index from Equation (\ref{bi}) may be $n<3$, which is consistent with observations. Hence, for pulsars with $n<3$, their internal fields are probably TD, rather than PD. \textbf{However, for the pulsar (PSR J1640-4631) with $n>3$, the results are relatively complicated. If $B_{\rm d}$ of PSR J1640-4631 follows the power-law decay form with ($a$, $\alpha$)=(0.01, 5/4) and (0.15, 5/4), the decay rates are small enough that no obvious differences can be found between the $\xi-\chi$ curves obtained by using $\dot{B}_{\rm d}$ calculated from Equation (\ref{dBdt}) and that obtained with $\dot{B}_{\rm d}\simeq0$, as one can see from Figure \ref{fig1}.\footnote{The results for ($a$, $\alpha$)=(0.01, 5/4) are not presented for brevity.} In these cases, without measurements of $\chi$ of PSR J1640-4631, no constraints can be set on its internal field configuration from its timing data.} The same conclusions were also obtained in \cite{2019PhRvD..99h3011C} and \cite{2023RAA....23e5020H}, which adopted an exponential form for the dipole field decay. Since the internal fields of PSR J1640-4631 may be either PD or TD, in panels (a) and (b) of Figure \ref{fig1} we show the curves of $\xi$ versus $\chi$ for the PD and TD cases, respectively. \textbf{In contrast, when $B_{\rm d}$ of PSR J1640-4631 decays following the power-law form with ($a$, $\alpha$)=(10, 1), its internal fields are probably TD because only in this configuration the $\xi-\chi$ curve can be obtained (see panel (b) of Figure \ref{fig1}).}  

\begin{table*}[h!]
	\centering
	\caption{The ranges of $\dot{B}_{\rm d}$ calculated by adopting different sets of values of ($a$, $\alpha$) for the young pulsars considered in this work.}
        \label{tab:tab2}
        \begin{threeparttable}
        \setlength{\tabcolsep}{1.2pt}
	\begin{tabular}{ccc} 
		\hline
		{Pulsar name} &\hspace{5em} {($a$,$\alpha$)} &\hspace{5em} {$\dot{B}_\mathrm{d}$ (G/s)}\\
		\hline
		PSR B0833-45 (Vela) &\hspace{5em} (0.01,5/4)&\hspace{5em} [$-5.80\times10^{-4}$,$-1.26\times10^{-3}$]\\
        {} &\hspace{5em} (0.15,5/4) &\hspace{5em} [$-8.72\times10^{-3}$,$-1.90\times10^{-2}$]\\
        {} &\hspace{5em} (10,1) &\hspace{5em} [$-7.02\times10^{-1}$,$-1.40\times10^{0}$]\\
		\hline
        PSR J1833-1034 &\hspace{5em} (0.01,5/4) &\hspace{5em} [$-6.60\times10^{-4}$,$-1.44\times10^{-3}$]\\
        {} &\hspace{5em} (0.15,5/4) &\hspace{5em} [$-9.96\times10^{-3}$,$-2.17\times10^{-2}$]\\
        {} &\hspace{5em} (10,1) &\hspace{5em} [$-7.90\times10^{-1}$,$-1.58\times10^{0}$]\\
		\hline
        PSR B0531+21(Crab) &\hspace{5em} (0.01,5/4) &\hspace{5em} [$-7.40\times10^{-4}$,$-1.62\times10^{-3}$]\\
        {} &\hspace{5em} (0.15,5/4) &\hspace{5em} [$-1.12\times10^{-2}$,$-2.44\times10^{-2}$]\\
        {} &\hspace{5em} (10,1) &\hspace{5em} [$-8.77\times10^{-1}$,$-1.75\times10^{0}$]\\
		\hline
        PSR J1513-5908 &\hspace{5em} (0.01,5/4) &\hspace{5em} [$-1.76\times10^{-2}$,$-3.84\times10^{-2}$]\\
        {} &\hspace{5em} (0.15,5/4) &\hspace{5em} [$-2.64\times10^{-1}$,$-5.77\times10^{-1}$]\\
        {} &\hspace{5em} (10,1) &\hspace{5em} [$-1.45\times10^{1}$,$-2.91\times10^{1}$]\\
		\hline
        PSR J1640-4631 &\hspace{5em} (0.01,5/4) &\hspace{5em} [$-1.51\times10^{-2}$,$-3.29\times10^{-2}$]\\
        {} &\hspace{5em} (0.15,5/4)&\hspace{5em} [$-2.26\times10^{-1}$,$-4.94\times10^{-1}$]\\
        {} &\hspace{5em} (10,1)&\hspace{5em}[$-1.27\times10^{1}$,$-2.53\times10^{1}$]\\
		\hline
	\end{tabular}
\end{threeparttable}
\end{table*}

\begin{figure*}[h!]
\centering
\begin{minipage}[t]{0.45\textwidth}
\includegraphics[width=3.2in]{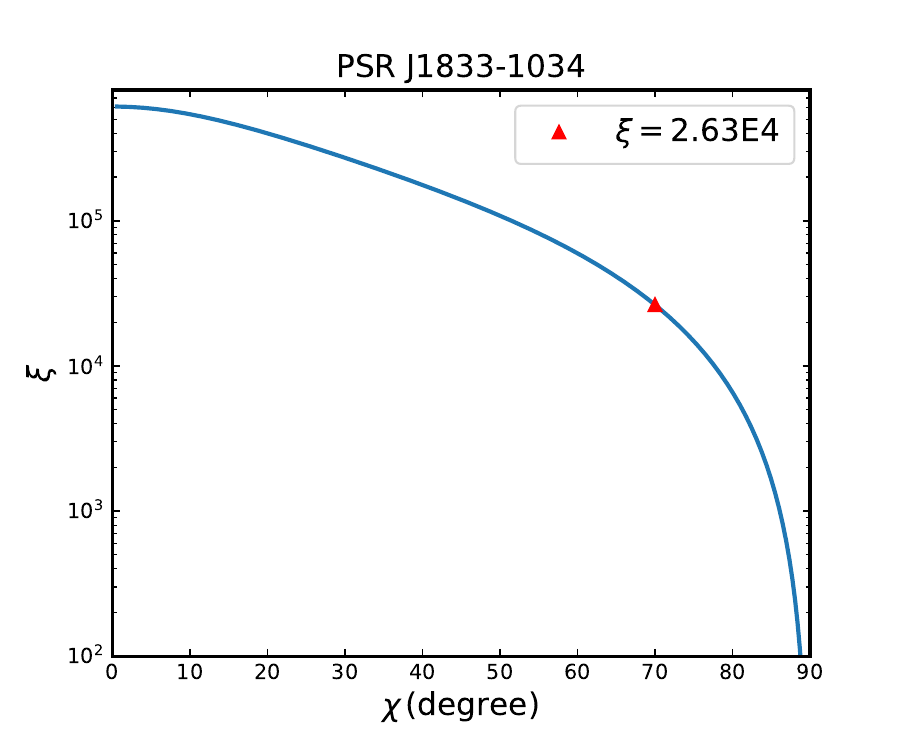}
\end{minipage}
\begin{minipage}[t]{0.45\textwidth}
\includegraphics[width=3.2in]{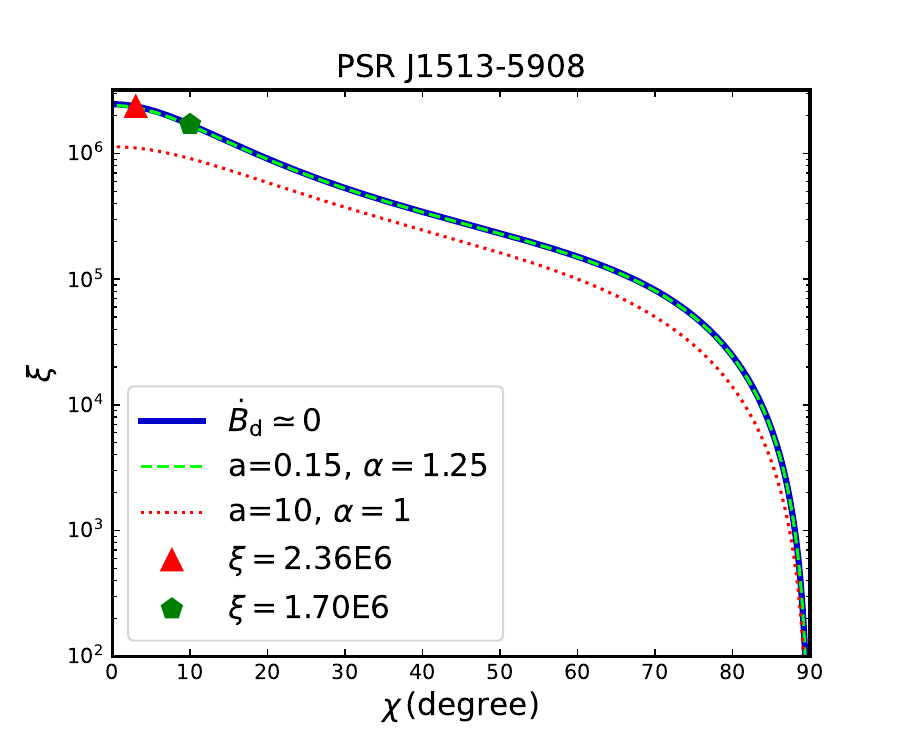}
\end{minipage}\\
\begin{minipage}[t]{0.45\textwidth}
\includegraphics[width=3.2in]{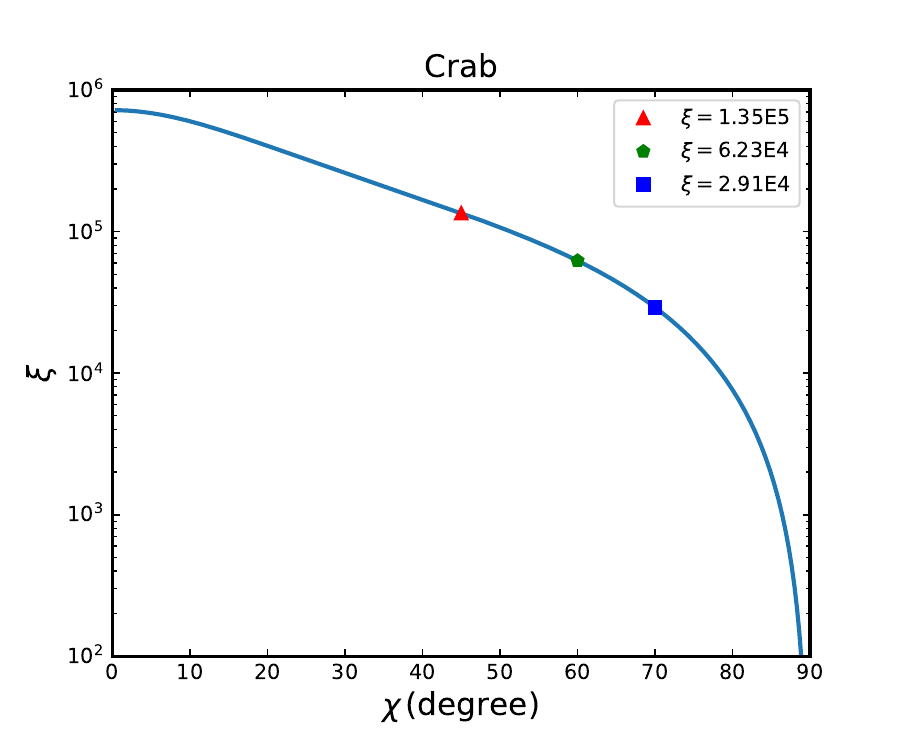}
\end{minipage}
\begin{minipage}[t]{0.45\textwidth}
\includegraphics[width=3.2in]{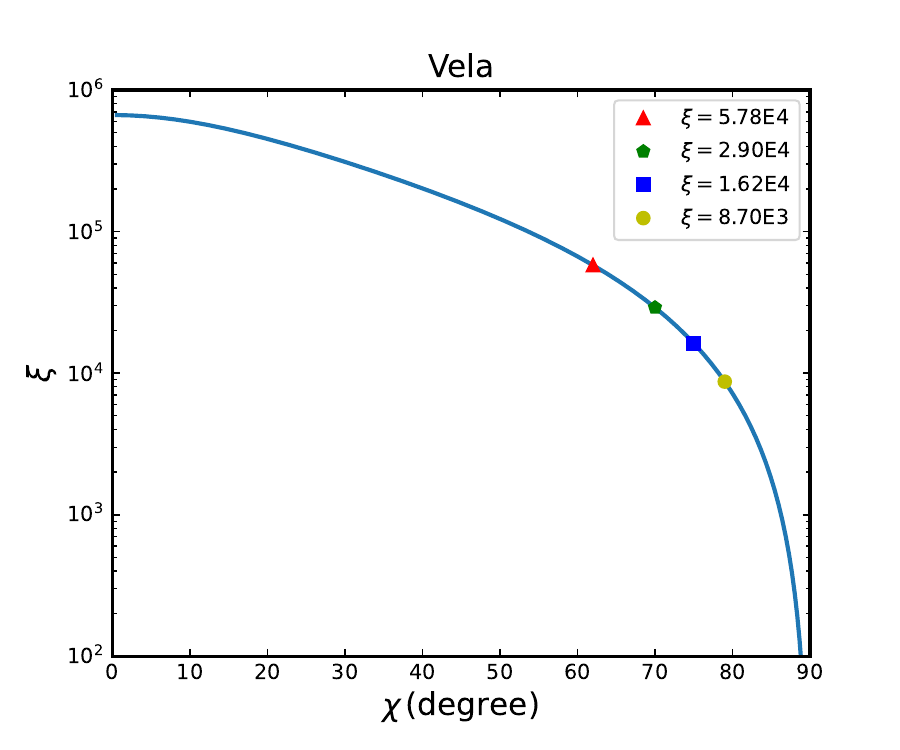}
\end{minipage}
\caption{The same as Figure \ref{fig1}, the results for PSR J1833-1034, PSR J1513-5908, the Crab and Vela pulsars are shown. All the results correspond to the case of TD internal fields. The colored points represent the values of $\xi$ at specific $\chi$ measured for these pulsars.
A comparison of the results obtained by using different decay rates $\dot{B}_{\rm d}$ is presented for PSR J1513-5908 (see the legends).}
\label{fig2}
\end{figure*}

\textbf{For PSR J1640-4631, constraints of $\xi$ can be found from Figure \ref{fig1}. If the field decay rate is small ($\dot{B}_{\rm d}\simeq0$), when PSR J1640-4631 has PD internal fields (panel (a)), its tilt angle should be within the ranges $0^{\circ}\lesssim \chi \lesssim18.5^{\circ}$ and $56^{\circ}\lesssim \chi \lesssim90^{\circ}$. Correspondingly, the number of precession cycles respectively distributes in the ranges $4\times10^3\lesssim \xi \lesssim2\times10^7$ and $5\times10^6\lesssim \xi \lesssim10^8$. In the case of small decay rate, when the internal fields of PSR J1640-4631 are TD (panel (b)), we have $18.5^{\circ}\lesssim \chi \lesssim56^{\circ}$ and $2\times10^6\lesssim \xi \lesssim10^8$. As we have shown above, in the case of large $\dot{B}_{\rm d}$, this pulsar probably has TD internal fields, and from panel (b) one can find $10^2\lesssim \xi \lesssim2.5\times10^6$. Thus the range of $\xi$ for PSR J1640-4631 derived here obviously differs from that obtained in \cite{2019PhRvD..99h3011C}.} Without measurements of $\chi$ of PSR J1640-4631, the value of $\xi$ also cannot be determined. For other young pulsars focused in this work, both their timing data and tilt angles can be measured, we therefore can set constraints on $\xi$. The results are presented in Figures \ref{fig2}. As these pulsars probably have TD internal fields, the $\xi-\chi$ curves thus correspond to the TD case. It is found that in the power-law decay scenario, the number of precession cycles is constrained to be in the range $10^4\lesssim\xi\lesssim{\rm a~few}\times10^6$, which is essentially the same as that derived based on the exponential decay scenario for these young pulsars \citep{2023RAA....23e5020H}. Moreover, the value of $\xi$ decreases with the increase of $\chi$ for the pulsars with $n<3$, as shown in Figures \ref{fig2}. \textbf{For PSR J1513-5908, we can also find that $\dot{B}_{\rm d}$ obtained through Equation (\ref{dBdt}) with ($a$, $\alpha$)=(0.15, 5/4) is so small that the resultant $\xi-\chi$ curve is the same as that obtained with $\dot{B}_{\rm d}\simeq0$. The larger $\dot{B}_{\rm d}$ derived with ($a$, $\alpha$)=(10, 1) leads to a relatively small change in the $\xi-\chi$ curve as compared to the case of $\dot{B}_{\rm d}\simeq0$.}
   
Here we pay special attention to the results of the Crab and Vela pulsars. From Figure \ref{fig2}, one can see that the measured tilt angles $\chi=45^\circ$, $60^\circ$, and $70^\circ$ of the Crab pulsar respectively correspond to $\xi=1.35\times10^5$, $6.23\times10^4$, and $2.91\times10^4$. The constraints on $\xi$ are consistent with the results\footnote{From modeling of the rise processes some large glitches of the Crab pulsar, the number of precession cycles is inferred to be within $3.3\times10^3\lesssim\xi\lesssim10^5$ \citep{2018MNRAS.481L.146H,2023RAA....23e5020H}.} obtained from observations of the glitch rise processes of the Crab pulsar \citep{2018MNRAS.481L.146H,2023RAA....23e5020H}. Similar conclusion is also obtained for the Vela pulsar. In the power-law decay scenario, from the measured angles $\chi=62^\circ$, $70^\circ$, $75^\circ$, and $79^\circ$ we respectively have $\xi=5.78\times10^4$, $2.90\times10^4$, $1.62\times10^4$, and $8.70\times10^3$, which are consistent with the range $\xi\lesssim1.8\times10^5$ derived from observations of the glitch rise processes of the Vela pulsar \citep{2019NatAs...3.1143A,2023RAA....23e5020H}. For the young pulsars whose $\chi$ can be measured, by substituting $\chi$ and the values of $\xi$ obtained above into Equation (\ref{chidot}) for $\epsilon_{\rm B}<0$, we can derive the theoretically predicted $\dot{\chi}$. The results are presented in Table \ref{tab3}. Consequently, if $\dot{\chi}$ of these pulsars can be measured from observations, then the power-law decay model of dipole field can be tested. From Table \ref{tab3} we can see that for the Crab pulsar, depending on the measured values of $\chi$, the calculated change rates are respectively $9.37\times10^{-12}$ rad/s, $12.6\times10^{-12}$ rad/s, and $18.3\times10^{-12}$ rad/s. These values will be compared to the $\dot{\chi}$ inferred from observations of the Crab pulsar and thus we can verify the validity of the power-law decay model.
\begin{table}
	\centering
	\caption{The measured values of tilt angles $\chi$ and theoretically predicted values of tilt angle change rates $\dot{\chi}$ of some young pulsars.}
        \label{tab3}
	\begin{tabular}{ccc} 
		\hline
		{Pulsar name} & {$\chi$} & {$\dot{\chi}$ ($10^{-12}$ rad/s)} \\
		\hline
		PSR B0833-45 (Vela) & $62^\circ$ & 4.82 \\
            {} & $70^\circ$ & 6.58 \\
            {} & $75^\circ$ & 8.68 \\
            {} & $79^\circ$ & 11.7 \\
		PSR J1833-1034 & $70^\circ$ & 10.9 \\
            PSR B0531+21 (Crab) & $45^\circ$ & 9.37 \\
            {} & $60^\circ$ & 12.6 \\
            {} & $70^\circ$ & 18.3 \\
            PSR 1513-5908 & $3^\circ$ & 15.6 \\
            {} & $10^\circ$ & 4.91 \\
		\hline
	\end{tabular}
\end{table}

\subsection{Testing the power-law decay model with observations of the Crab pulsar}

Till now, the Crab pulsar is the only pulsar whose $\dot{\chi}$ can be inferred from observations \citep{2013Sci...342..598L}. This pulsar thus represents the only sample on which the power-law decay model can be tested. \cite{2013Sci...342..598L} reported that the separation between the main pulse and interpulse of the Crab pulsar increases steadily at a rate $0.62^{\circ}\pm0.03^{\circ}$ per century, which can be accounted for if its tilt angle increases at a comparable rate. Therefore, the inferred tilt angle change rate is $\dot{\chi}=3.43\times 10^{-12}$ rad/s by neglecting the error bars. The inferred $\dot{\chi}>0$ supports our suggestion that the internal fields of the Crab pulsar may be TD. However, the inferred $\dot{\chi}$ is about several times smaller than the theoretical values predicted by the power-law decay model, as one can see from Table \ref{tab3}. This means that the power-law decay model probably cannot be applied to the discussion of dipole field evolution of the Crab pulsar. 

\begin{figure}[h!]
\centering
 \includegraphics[scale=0.5]{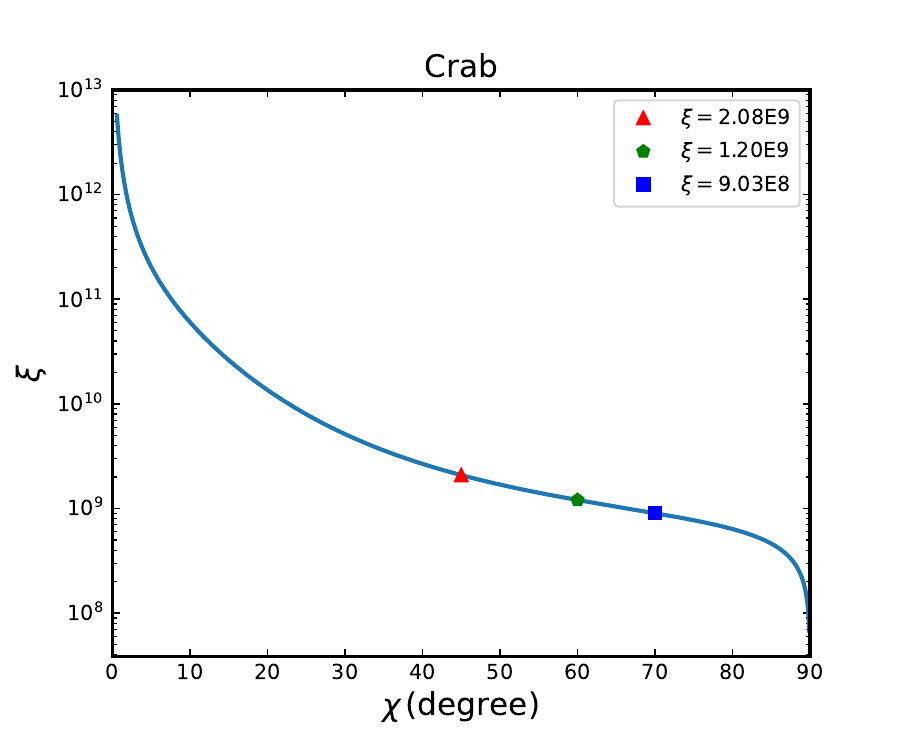}
 \caption{The curve of $\xi$ versus $\chi$ obtained by using the $\dot{\chi}$ inferred from observations and the timing data of the Crab pulsar. The colored points represent the values of $\xi$ at specific $\chi$ measured.}
 \label{fig3}
\end{figure}

The contradiction between the power-law decay model and the Crab's observations can also be found in Figure \ref{fig3}, which show the evolution of $\xi$ versus $\chi$. Different from Figure \ref{fig2}, the $\xi-\chi$ curve in this figure is obtained by substituting the inferred $\dot{\chi}$ and $P$ into Equation (\ref{chidot}) for $\epsilon_{\rm B}<0$. From Figure \ref{fig3}, we can see that the observed $\chi=45^\circ$, $60^\circ$, and $70^\circ$ respectively correspond to $\xi=2.08 \times 10^9$, $1.2 \times 10^9$, and $9.03 \times 10^8$, which are all much larger than the upper limit $\xi_{\rm upp}\approx10^8$ when phonon excitations govern the vortex-lattice interaction \citep{2018ASSL..457..401H,2018MNRAS.481L.146H,2019PhRvD..99h3011C,2023RAA....23e5020H}. Hence, to satisfy both the Crab’s timing data and the $\dot{\chi}$ inferred from observations, an unreasonable large $\xi\sim 10^9$ is required. This represents another reason why the power-law decay model cannot be applied to the Crab pulsar.

\begin{figure}[h!]
\centering
 \includegraphics[scale=0.5]{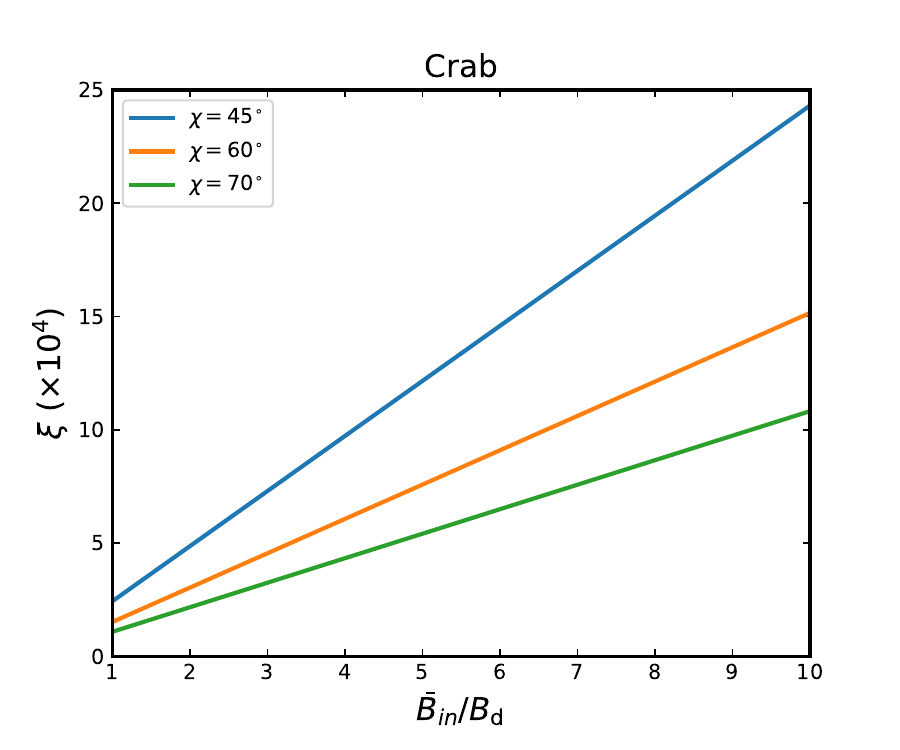}
 \caption{Evolution of $\xi$ versus the strength ratio of internal toroidal field to surface dipole field $\bar{B}_{\rm in}/B_{\rm d}$. The colored lines represent the $\xi-\bar{B}_{\rm in}/B_{\rm d}$ curves calculated at specific $\chi$ measured for the Crab pulsar, as shown in the legends.}
 \label{fig4}
\end{figure}

\begin{figure}[h!]
\centering
 \includegraphics[scale=0.5]{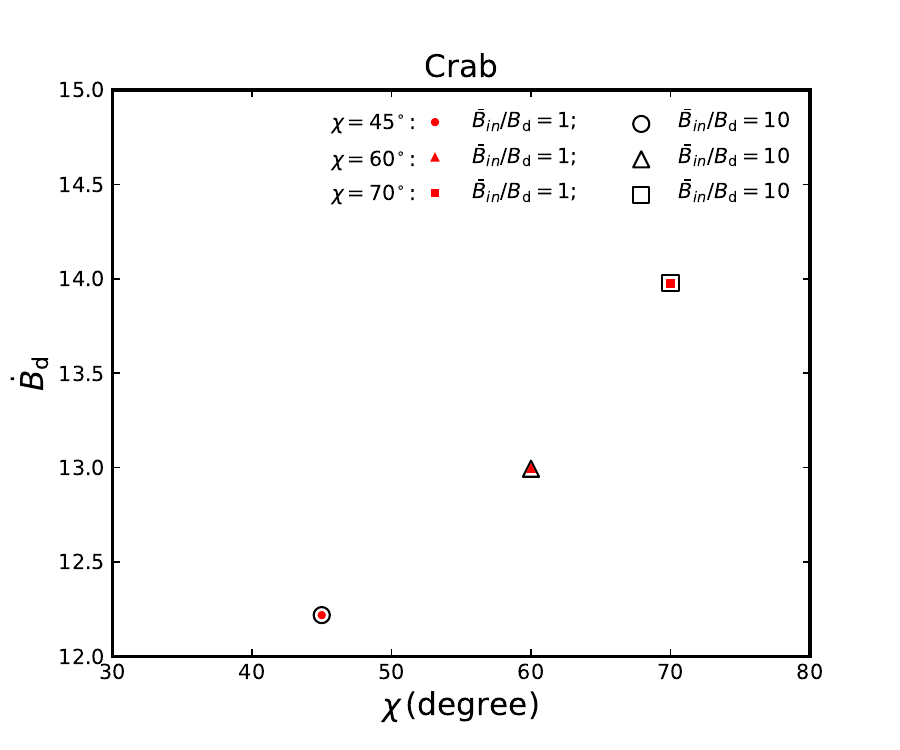}
 \caption{The dipole field change rates $\dot{B_{\rm d}}$ calculated at specific $\chi$ measured and by assuming different values of $\bar{B}_{\rm in}/B_{\rm d}$ for the Crab pulsar, as shown in the legends.}
 \label{fig5}
\end{figure}

It is thus interesting to investigate what evolution behavior the dipole field should follow in order to account for the measured $P$, $\dot{P}$, $n$, $\chi$, and inferred $\dot{\chi}$ of the Crab pulsar. We find that to properly account for the Crab's observational data, rather than decay, the dipole field may increase with time. The method is as follows. After substituting $\chi$, $\dot{\chi}$, $P$, and Equation (\ref{Bd}) into Equation (\ref{chidot}) for $\epsilon_{\rm B}<0$, and taking the strength ratio of internal toroidal field to surface dipole field in the reasonable range $1\leq\bar{B}_{\rm in}/B_{\rm d}\leq10$ \citep{2012PhRvL.109h1103G,2019PhRvD..99h3011C}, by solving this equation we can get the curves of $\xi$ versus $\bar{B}_{\rm in}/B_{\rm d}$ for different values of $\chi$, as presented in Figure \ref{fig4}. Regardless of the measured $\chi$, the value of $\xi$ increases monotonically with the increase of $\bar{B}_{\rm in}/B_{\rm d}$, and distributes in the range $\sim10^4-10^5$. We note that such a range is consistent with the result obtained from modeling of the glitch rise processes of the Crab pulsar \citep{2018MNRAS.481L.146H}. We then substitute the observed $P$, $\dot{P}$, $n$, $\chi$, inferred $\dot{\chi}$, and Equation (\ref{Bd}) into Equation (\ref{bi}), and take $\bar{B}_{\rm in}/B_{\rm d}=1$ and 10, respectively, the values of $\dot{B_{\rm d}}$ at each measured $\chi$ can be calculated. The results are shown in Figure \ref{fig5}. Irrespective of the strength ratio is $\bar{B}_{\rm in}/B_{\rm d}=1$ or 10, the dipole field change rate $\dot{B_{\rm d}}$ remains the same at a specific $\chi$ measured. For $\chi=45^\circ$, $60^\circ$, and $70^\circ$, the field change rate generally distributes in the range $\dot{B_{\rm d}}\sim12-14$ G/s, which suggests that instead of decaying with time, $B_{\rm d}$ should increase with time, in order to account for all of the observations of the Crab pulsar. Increase of the dipole field of the Crab pulsar may be caused by reemergence of the initial dipole field that was submerged into the NS crust by the accreted matter after the birth of this NS \citep{1999A&A...345..847G,2012ApJ...748..148S}.

\section{Conclusion and discussions} \label{conclusion}
Although the dipole fields of NSs are generally considered to decay with time due to either Hall drift and Ohmic dissipation if they originate from the NS crust, or ambipolar diffusion if they follow the decay of the core fields, the mathematical form of dipole field decay is still an open issue. The mathematical form is not only important for quantitative studies of the evolution of NSs (e.g., \citealt{1998ApJ...506L..61H,2000ApJ...529L..29C,2012MNRAS.422.2878D,2016ApJ...833..261B}), but also quite critical for the understanding of specific field decay mechanisms in NSs \citep{1992ApJ...395..250G}. In this work, we adopt the power-law form of field decay proposed in \cite{2000ApJ...529L..29C} to depict the evolution of $B_{\rm d}$, and use the observed timing data and $\chi$ of several young pulsars with a steadily measured $n$ and an ordinary dipole field $B_{\rm d}\sim10^{12}-10^{13}$ G to set constraints on the internal field configurations and the number of precession cycles $\xi$ of these NSs. Our results show that for the young pulsars with $n<3$, their internal fields are probably TD, and the number of precession cycles are generally within $10^4\lesssim\xi\lesssim{\rm a~few}\times10^6$. \textbf{However, for PSR J1640-4631 (which has $n>3$), relatively complicated results are found. If $B_{\rm d}$ of this pulsar decays following the power-law form with ($a$, $\alpha$)=(0.01, 5/4) and (0.15, 5/4), the decay rates will be small, and in these cases neither its internal field configuration nor the value of $\xi$ can be determined without measurements of its $\chi$. In contrast, if the power-law form with ($a$, $\alpha$)=(10, 1) is adopted, then the decay rate will be relatively large, and in this case its internal fields are probably TD and the number of precession cycles is within $10^2\lesssim \xi \lesssim2.5\times10^6$.} Thus the range of $\xi$ for PSR J1640-4631 derived here obviously differs from that obtained in \cite{2019PhRvD..99h3011C}. For other young pulsars, the constraints on $\xi$ are essentially the same as that derived based on the exponential decay of dipole field \citep{2019PhRvD..99h3011C,2023RAA....23e5020H}. Especially, for the Crab and Vela pulsars, the values of $\xi$ obtained in this work are consistent with the constraints derived from observations of the glitch rise processes of the two pulsars \citep{2018MNRAS.481L.146H,2019NatAs...3.1143A,2023RAA....23e5020H}. 

We note that whether the power-law form of field decay can be applied to NSs with dipole fields of ordinary strength still remains unknown. Combining the inferred $\dot{\chi}$ of the Crab pulsar \citep{2013Sci...342..598L} with our theoretical model, this issue may be verified. Our results show that if $B_{\rm d}$ of the Crab pulsar follows the power-law decay, $\dot{\chi}$ predicted by the model would be several times larger than the inferred one. Moreover, in the power-law decay scenario, in order to satisfy the Crab’s timing data and the inferred $\dot{\chi}$, a rather large $\xi\sim10^9$ is required, which may be physically unreasonable. We conclude that the power-law decay model is inconsistent with the observational data of the Crab pulsar and thus cannot be applied to this pulsar. However, for other young pulsars focused in the paper, it is still unknown whether the power-law decay model is applicable since their $\dot{\chi}$ are currently not available from observations. Future measurements of $\dot{\chi}$ of these pulsars will eventually help to ascertain whether the power-law decay model can be applied to NSs with ordinary fields. Interesting, we find that to properly account for all of the Crab's observations, rather than decay with time, the Crab's dipole field should increase with time at a rate $\dot{B_{\rm d}}\sim12-14$ G/s for a reasonable range of the strength ratio $1\leq\bar{B}_{\rm in}/B_{\rm d}\leq10$ \citep{2012PhRvL.109h1103G,2019PhRvD..99h3011C}. Meanwhile, the number of precession cycles is constrained to be within $\xi\sim10^4-10^5$, consistent with the results obtained from modeling of the rise processes of some large glitches of the Crab pulsar \citep{2018MNRAS.481L.146H,2023RAA....23e5020H}. 

For the Crab pulsar, the unavailability of the power-law decay model may indicate that such a mathematical form could only be applied to NSs with $B_{\rm d}>10^{13}$ G, as suggested in \cite{2000ApJ...529L..29C}. In other words, for NSs with dipole fields of ordinary strength, the dominant field decay mechanisms may be neither Hall drift nor ambipolar diffusion. A definite conclusion may be drawn if $\dot{\chi}$ of other young pulsars can be measured. Another possibility is that at current stage, rather than decay due to Hall drift or ambipolar diffusion, the evolution of the Crab's dipole field may mainly manifest as reemergence of the field that submerged into the crust because of fall-back accretion after the birth of the NS \citep{1999A&A...345..847G,2012ApJ...748..148S}. In future work, based on the field reemergence model, we will estimate the initial dipole field, the mass of accreted matter that needed to submerge the initial dipole field, the number of precession cycles, and the internal field configuration of these young pulsars. 

\textbf{Though the power-law decay form of dipole field seems to be disfavored by the Crab's observations, it is still interesting to discuss whether other decay forms could be compatible with the observations. If $B_{\rm d}$ of the Crab pulsar follows the exponential decay as given in Equation (6) of \cite{2019PhRvD..99h3011C}, the decay rate $\dot{B}_{\rm d}$ will be determined by the decay time scale $\tau_{\rm D}$, whose value is possibly larger than a few$\times10^5$ yrs, as shown in Figure 6 of \cite{2023RAA....23e5020H}. In this case, the resultant decay rate satisfies $\dot{B}_{\rm d}\gtrsim-0.6$ G/s for the Crab pulsar, we can thus approximately take $\dot{B}_{\rm d}\simeq0$, which is the same as the results of power-law decay. It means that the exponential decay model is also disfavored by observations of the Crab pulsar, and an increasing $B_{\rm d}$ as proposed above is required to account for the observations. Another simple form of field decay is the nonlinear form as shown in Equation (7) of \cite{2019PhRvD..99h3011C}. Since the Crab's age is much smaller than $\tau_{\rm D}$, the nonlinear form is thus essentially the same as the exponential decay form, and also inconsistent with the Crab's observations. In future work, other field decay models with more complicated forms will be tested with observations of the Crab pulsar, and also other young pulsars if more observational data is available.} 

\section*{Declaration of competing interest}
The authors declare that they have no known competing financial interests or personal relationships that could have appeared to influence the work reported in this paper.

\section*{Data availability}
Observational data used in this paper are quoted from the cited works. Additional data generated from computations will be shared on reasonable
request to the corresponding authors.

\section*{Acknowledgements}
We gratefully thank the anonymous referee for helpful comments which helped improving the manuscript. This work is supported by the National Natural Science Foundation of China (Grant Nos. 12003009, and 12033001), CAS “Light of West China” Program (Grant No. 2019-XBQNXZ-B-016), and the National SKA program of China (Grant No. 2020SKA0120300).

\end{document}